\documentclass[footinbib,aps,prb,twocolumn,superscriptaddress,groupedaddress,a4paper]{revtex4}  
\usepackage{graphicx}
\usepackage{dcolumn}
\usepackage{bm}
\usepackage{amssymb}
\usepackage{amsmath}
\usepackage{subfigure}
\usepackage{color}
\usepackage{lscape}
\begin{document}

\widetext
\title{First principles analysis of electronic structure evolution and the\\indirect- to direct-gap transition in Ge$_{1-x}$Pb$_{x}$ group-IV alloys}


\author{Christopher A.~Broderick}
\email{c.broderick@umail.ucc.ie} 
\affiliation{Tyndall National Institute, University College Cork, Lee Maltings, Dyke Parade, Cork T12 R5CP, Ireland}
\affiliation{Department of Physics, University College Cork, Cork T12 YN60, Ireland}

\author{Edmond J.~O'Halloran}
\affiliation{Tyndall National Institute, University College Cork, Lee Maltings, Dyke Parade, Cork T12 R5CP, Ireland}
\affiliation{School of Chemistry, University College Cork, Cork T12 YN60, Ireland}

\author{Eoin P.~O'Reilly}
\affiliation{Tyndall National Institute, University College Cork, Lee Maltings, Dyke Parade, Cork T12 R5CP, Ireland}
\affiliation{Department of Physics, University College Cork, Cork T12 YN60, Ireland}

\date{\today}


\begin{abstract}

We present a theoretical analysis of electronic structure evolution in the group-IV alloy Ge$_{1-x}$Pb$_{x}$ based on density functional theory. For ordered alloy supercells we demonstrate the emergence of a singlet conduction band (CB) edge state, suggesting the emergence of a direct band gap for Pb compositions as low as $x \approx 1$\%. However, application of hydrostatic pressure reveals Pb-induced hybridisation, with the CB edge state in a Ge$_{63}$Pb$_{1}$ ($x = 1.56$\%) supercell retaining primarily indirect (Ge L$_{6c}$) character. For an ordered Ge$_{15}$Pb$_{1}$ ($x = 6.25$\%) supercell we find that the CB edge has acquired primarily direct (Ge $\Gamma_{7c}$) character, confirming the presence of an indirect- to direct-gap transition. The importance of alloy disorder in determining the details of the electronic structure is highlighted by investigating the formation of a nearest-neighbour Pb-Pb pair, which produces a large decrease (increase) in band gap (spin-orbit splitting energy) compared to equivalent structures having large Pb-Pb separation. Having established the importance of short-range disorder, we analyse the electronic structure evolution as a function of $x$ using a series of 128-atom special quasi-random structures (SQSs). Our calculations reveal a strong reduction (increase) of the band gap (spin-orbit splitting energy), by $\approx 100$ meV ($\approx 40$ meV) per \% Pb replacing Ge. We find an indirect- to direct-gap transition occurring in a narrow composition range centred about $x \approx 7$\%, close to which composition we calculate that the alloy becomes semimetallic. Further analysis suggests that long-range order introduced by Born von Karman (supercell) boundary conditions leads to overestimated energy splitting of the Ge L$_{6c}$-derived CB states in the 128-atom SQSs. Accounting for these finite-size effects, we expect a direct band gap to emerge in Ge$_{1-x}$Pb$_{x}$ for $x \gtrsim 3$ -- 4\%.

\end{abstract}


\maketitle


\section{Introduction}
\label{sec:introduction}


Despite their prominence in contemporary microelectronics, the group-IV semiconductors silicon (Si) and germanium (Ge) are unsuitable for applications in active photonic devices due to their indirect band gaps, which make them intrinsically inefficient emitters and absorbers of light. Given the strong impetus to develop novel hybrid optoelectronic architectures, significant research effort has been dedicated over the past two decades to Si photonics. \cite{Zhou_LSAA_2015,Wang_LPR_2017,Shen_JLT_2019} The key aim of Si photonics is the development of photonic components which are compatible with established complementary metal-oxide semiconductor (CMOS) processing infrastructure, to deliver step-changes in device performance and capabilities, either via on-chip integration with microelectronics or by facilitating optical interconnection between optoelectronic chips. \cite{Sun_Nature_2015,Atabaki_Nature_2018} While significant progress has been made in the development of passive photonic components such as waveguides and modulators, the development of Si photonics is currently limited by the lack of direct-gap materials suitable for application as CMOS-compatible semiconductor lasers and light-emitting diodes. \cite{Thomson_JO_2016,Soref_OPN_2016}

Due to the small difference of $\approx 150$ meV between the indirect (fundamental) L$_{6c}$-$\Gamma_{8v}$ and direct $\Gamma_{7c}$-$\Gamma_{8v}$ band gaps of Ge, there has recently been a strong surge of interest in engineering the conduction band (CB) structure of Ge via strain, doping or alloying in order to bring about a direct fundamental band gap. \cite{Sun_APL_2009,Zhang_PRL_2009,Liu_OL_2009,Gupta_JAP_2013,Geiger_FM_2015,Saito_SST_2016,Reboud_PCGC_2017} Much recent attention has focused on Ge$_{1-x}$Sn$_{x}$ alloys, where it has been predicted that incorporation of 6 -- 9\% Sn is sufficient to bring about a direct band gap. \cite{Kouvetakis_ARMR_2006,Soref_PTRSA_2014,Zaima_STAM_2015} Experimental confirmation of the emergence of a direct band gap in Ge$_{1-x}$Sn$_{x}$ -- culminating in initial demonstrations of optically pumped lasing \cite{Wirths_NP_2015,Al-Kabi_APL_2016,Reboud_APL_2017,Margetis_ACSP_2017,Margetis_APL_2018,Zhou_ACSP_2019} -- has stimulated broader interest in related group-IV alloys containing carbon (C) \cite{Stephenson_JAP_2016,Stephenson_JEM_2016,Kirwan_SST_2019,Broderick_JAP_2019} or lead (Pb). \cite{Huang_PB_2014,Huang_JAC_2017,Broderick_NUSOD_2019}


While Ge$_{1-x}$Sn$_{x}$ alloys have attracted significant theoretical interest, there have been few reports to date regarding Ge$_{1-x}$Pb$_{x}$ alloys. Initial density functional theory (DFT) calculations by Huang et al.~\cite{Huang_PB_2014} predicted (i) a strong band gap reduction in response to Pb incorporation, (ii) that the indirect- to direct-gap transition occurs for Pb compositions as low as $x \approx 1$\%, and (iii) that the alloy band gap closes for $x \approx 2$\%. However, these calculations possess several shortcomings. Firstly, the calculated conduction band (CB) minimum of Ge in Ref.~\onlinecite{Huang_PB_2014} lies along the $\Delta$ direction in the Brillouin zone, rather than at the L point. Secondly, the use of a local density approximation (LDA) exchange-correlation (XC) functional leads to band gap underestimation. More recently, the same authors presented calculations based on a more accurate GGA+U XC functional, and revised the estimated Pb composition for the transition to a direct gap to $x \approx 3.5$\%. \cite{Huang_JAC_2017} However, little work has yet been undertaken to analyse the nature of the indirect- to direct-gap transition, its consequences for the alloy electronic structure, and hence its implications for device applications.


Given the recent establishment of epitaxial growth of Ge$_{1-x}$Pb$_{x}$ alloys,  \cite{Alahmad_JEM_2018,Liu_JAC_2019} and initial experimental evidence for the emergence of a direct band gap, \cite{Alahmad_JEM_2018} detailed theoretical insight into the alloy electronic structure is required to guide the development of suitable materials for potential device applications. In this paper, we present a theoretical analysis of electronic structure evolution in Ge$_{1-x}$Pb$_{x}$ alloys, using first principles calculations based on DFT. Our alloy supercell calculations firstly demonstrate that Pb incorporation strongly impacts the CB structure. We note that this is contrary to expectations based on conventional chemical trends, on which basis the large covalent radius and reduced electronegativity of Pb compared to Ge would suggest that Pb incorporation primarily impacts the VB structure (akin, e.g., to Bi incorporation in GaAs). Our analysis further reveals a strong Pb-induced reduction of the fundamental band gap and increase of the VB spin-orbit splitting energy, as well as a strong sensitivity of the alloy CB structure to the presence of short-range alloy disorder (Pb clustering). From a theoretical perspective, the presence of such alloying effects indicates a breakdown of the virtual crystal approximation (VCA), upon which previous analysis \cite{Huang_PB_2014} of Ge$_{1-x}$Pb$_{x}$ alloys has been based.

Due to similarities in the impact of Pb and Sn incorporation on the Ge band structure, we discuss our results via comparison to the results of equivalent calculations for Ge$_{1-x}$Sn$_{x}$ alloys. \cite{Broderick_NUSOD_2019} We demonstrate that the electronic structure evolution in Ge$_{1-x}$Pb$_{x}$ admits important quantitative differences compared to that in Ge$_{1-x}$Sn$_{x}$. Confirming the strong impact of Pb incorporation on the band structure, we firstly calculate that Pb incorporation results in a strong reduction of the fundamental band gap, by $\approx 100$ meV per \% Pb replacing Ge. Secondly, the band gap becomes direct in character with increasing $x$, but the alloy CB edge eigenstates in our supercell calculations are in general neither purely indirect nor direct in character, but predominantly contain an admixture of indirect (Ge L$_{6c}$) and direct (Ge $\Gamma_{7c}$) character. Via analysis of the electronic structure evolution with Pb composition $x$ we identify a relatively abrupt indirect- to direct-gap transition, occurring in a narrow composition range centred about $x \approx 7$\% in 128-atom special quasi-random structures (SQSs), close to the composition at which we calculate that Ge$_{1-x}$Pb$_{x}$ becomes semimetallic. We note however that finite-size effects -- associated with the introduction of long-range order due to the use of Born von Karman (supercell) boundary conditions -- likely leads to overestimation of the Pb composition at which the indirect- to direct-gap transition is predicted to occur. Based on qualitative consideration of these finite-size effects, we estimate the direct gap to shift below the indirect gap in energy for Pb compositions $x \approx 3$ -- 4\%.


The remainder of this paper is organised as follows. In Sec.~\ref{sec:theoretical_model} we describe the first principles framework used to calculate the structural and electronic properties of Ge$_{1-x}$Pb$_{x}$ alloy supercells, as well as the generation of the SQSs used in our analysis of disordered alloys. We also describe in this section how we use calculations of the band structure as a function of hydrostatic pressure to determine the character of the alloy band gap. The results of our calculations are presented in Sec.~\ref{sec:results}, beginning in Sec.~\ref{sec:results_ordered} with an analysis of the impact of Pb incorporation on the electronic structure of ordered Ge$_{1-x}$Pb$_{x}$ alloy supercells. In Sec.~\ref{sec:results_clustering} we quantify the importance of the local environment around a substitutional Pb atom, via analysis of the impact of nearest-neighbour Pb-Pb pair formation on the electronic structure. Then, in Sec.~\ref{sec:results_disordered} we analyse the evolution of the electronic structure with $x$ in 128-atom disordered (SQS) alloy supercells. Finally, in Sec.~\ref{sec:conclusions} we summarise and conclude.


\section{Theoretical model}
\label{sec:theoretical_model}

Our analysis of the Ge$_{1-x}$Pb$_{x}$ electronic structure is based on DFT calculations employing two distinct exchange-correlation (XC) functionals: (i) the Heyd-Scuseria-Ernzerhof hybrid XC functional \cite{Heyd_JCP_2003,Heyd_JCP_2004} modified for solids (HSEsol), \cite{Schimka_JCP_2011} and (ii) the Tran-Blaha modified Becke-Johnson (mBJ) XC functional. \cite{Tran_PRL_2009} The semi-core $d$ states of Ge are treated as core electron states, since unfreezing these states has been demonstrated to have a negligible impact on the calculated electronic structure. \cite{Eckhardt_PRB_2014} We adopt the same approximation for Pb. We therefore employ pseudopotentials in which the (4$s$)$^{2}$(4$p$)$^{2}$ orbitals of Ge and (6$s$)$^{2}$(6$p$)$^{2}$ orbitals of Pb are explicitly treated as valence states. Due to the large relativistic effects associated with Pb, and to a lesser extent with Ge, all calculations include spin-orbit coupling. All DFT calculations were performed using the projector augmented-wave method, \cite{Blochl_PRB_1994,Kresse_PRB_1999} as implemented in the Vienna Ab-initio Simulation Package (VASP). \cite{Kresse_PRB_1996,Kresse_CMS_1996}


\begin{table*}[t!]
	\caption{\label{tab:dft_benchmark} Lattice constant $a$, direct $\Gamma_{7c}$-$\Gamma_{8v}$ band gap $E_{g}$ and $\Gamma_{8v}$-$\Gamma_{7v}$ VB spin-orbit splitting energy $\Delta_{\protect\scalebox{0.6}{\textrm{SO}}}$ for Ge, diamond-structured Pb (d-Pb) and zinc blende GePb (zb-GePb), calculated via DFT using the HSEsol (with $\alpha = 0.3$), and LDA (for $a$) or mBJ (with $c = 1.2$, for $E_{g}$ and $\Delta_{\protect\scalebox{0.6}{\textrm{SO}}}$) XC functionals, and compared to low-temperature experimental measurements and previous first principles theoretical calculations. For Ge the fundamental (indirect) L$_{6c}$-$\Gamma_{8v}$ band gap is listed in parentheses.}
	\begin{ruledtabular}
		\begin{tabular}{c|ccc|ccc|ccc}
			            &        & $a$ (\AA) &                          &               & $E_{g}$ (eV)  &                     &        & $\Delta_{\scalebox{0.6}{\textrm{SO}}}$ (eV) &             \\
			Material    & HSEsol & LDA       & Reference                & HSEsol        & mBJ           & Reference           & HSEsol & mBJ                                         & Reference   \\
			\hline
			Ge          & 5.653  & 5.649     & 5.657$^{a}$, 5.648$^{b}$ & 0.908 (0.765) & 0.868 (0.724) & 0.890 (0.744)$^{c}$ & 0.322  & 0.274                                       & 0.296$^{c}$ \\
			d-Pb        & 6.907  & 6.852     & 6.673$^{d}$, 7.074$^{e}$ & $-$5.329      & $-$4.605      & -----               & 2.422  & 1.999                                       & 2.377$^{f}$ \\
			zb-GePb     & 6.317  & 6.297     & 6.154$^{d}$, 6.265$^{g}$ & $-$2.695      & $-$2.407      & $-$2.250$^{g}$      & 1.123  & 1.029                                       & 0.930$^{g}$ \\
		\end{tabular}
	\end{ruledtabular}
	\begin{flushleft}
	$^{a}$Meas.~(avg.), Ref.~\onlinecite{Landolt_1982_1} \;
	$^{b}$Calc.~(avg.), Ref.~\onlinecite{Landolt_1982_1} \;
	$^{c}$Meas., Ref.~\onlinecite{Landolt_1982_2} \;
	$^{d}$Calc., Ref.~\onlinecite{Wang_PRB_2002} \;
	$^{e}$Calc.~(avg.), Ref.~\onlinecite{Hermann_PRB_2010} \;
	$^{f}$Calc., Ref.~\onlinecite{Herman_PRL_1963} \;
	$^{g}$Calc., Ref.~\onlinecite{Hammou_PSSC_2017} \;
	\end{flushleft}
\end{table*}

Our DFT calculations are based closely on those we have established recently for Ge$_{1-x}$Sn$_{x}$ alloys, full details of which can be found in Ref.~\onlinecite{Halloran_OQE_2019}. Since Pb incorporation is found to strongly impact the Ge band structure close in energy to the CB edge, \cite{Broderick_NUSOD_2019} and since we are interested in the transition from an indirect to direct band gap in Ge$_{1-x}$Pb$_{x}$ at low Pb compositions $x \lesssim 10$\%, we prioritise the accuracy of the description of the Ge band structure close in energy to the CB edge. For the HSEsol (mBJ) calculations we therefore treat the exact exchange mixing $\alpha$ (Becke-Roussel mixing parameter $c$) in the XC functional as an adjustable parameter, the value of which is chosen to reproduce the measured difference in energy between the fundamental indirect L$_{6c}$-$\Gamma_{8v}$ and direct $\Gamma_{7c}$-$\Gamma_{8v}$ band gaps of Ge -- i.e.~the separation in energy between the L$_{6c}$ and $\Gamma_{7c}$ CB edge states of Ge. We respectively find that setting $\alpha = 0.3$ and $c = 1.2$ in the HSEsol and mBJ XC functionals accurately reproduces the measured 146 meV $\Gamma_{7c}$-L$_{6c}$ energy difference. \cite{Halloran_OQE_2019,Landolt_1982_2} For the HSEsol calculations we use a screening (range separation) parameter $\mu = 0.2$ \AA$^{-1}$. \cite{Heyd_JCP_2003,Heyd_JCP_2004,Schimka_JCP_2011}

For primitive (2-atom diamond) unit cells we utilise a $\Gamma$-centred $8 \times 8 \times 8$ Monkhorst-Pack \textbf{k}-point grid for Brillouin zone integration, which is downsampled for larger supercells to preserve the resolution of the Brillouin zone sampling. A plane wave cut-off energy of 400 eV is used for all calculations, which is chosen to be sufficiently high to minimise Pulay stress and allow for accurate structural relaxation. Structural relaxation is achieved via free energy minimisation, by allowing the lattice vectors and ionic positions to relax freely, subject to the additional stopping criterion that the maximum force on any atom in the supercell does not exceed 0.01 eV \AA$^{-1}$. To generate relaxed atomic positions for HSEsol (mBJ) electronic structure calculations, the HSEsol (LDA) XC functional is used to perform structural relaxation. Since relativistic effects in Pb are sufficiently large to impact structural properties -- e.g.~calculations neglecting spin-orbit coupling incorrectly predict that elemental Pb is diamond-structured rather than FCC-structured in equilibrium \cite{Christensen_PRB_1986} -- spin-orbit coupling is explicitly included for all structural relaxations.

\begin{figure*}[t!]
	\includegraphics[width=1.00\textwidth]{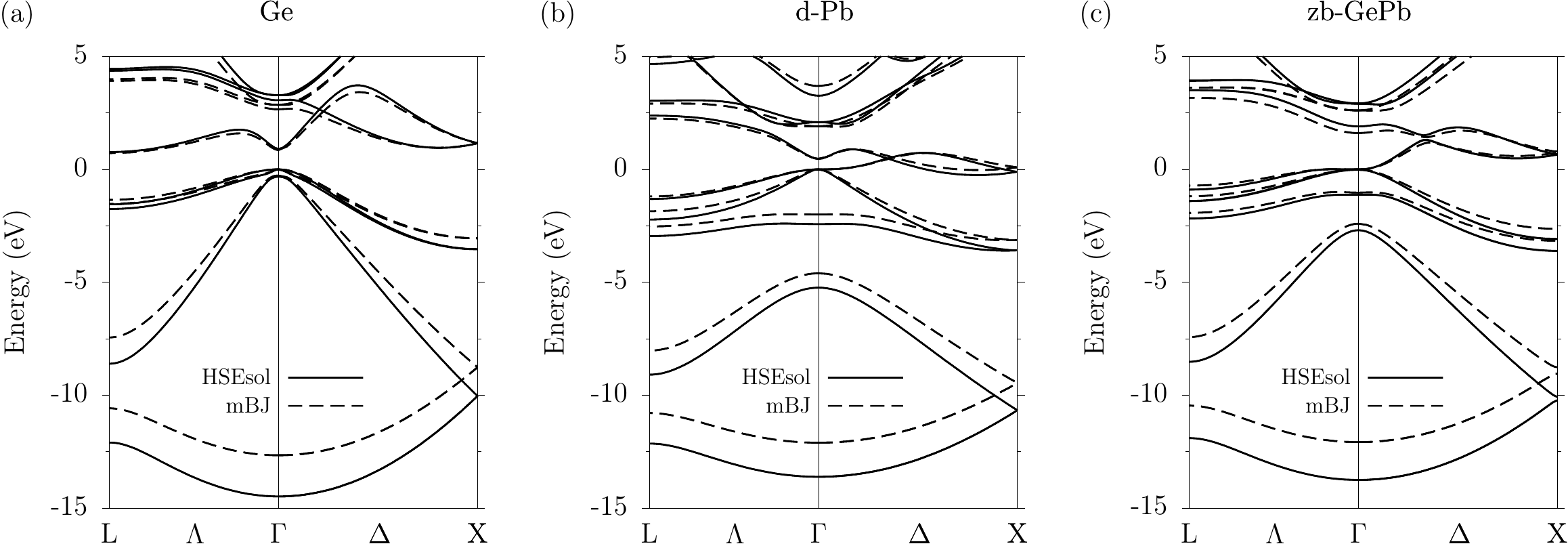}
	\caption{Band structure of (a) Ge, (b) diamond-structured Pb (d-Pb), and (c) zinc blende GePb (zb-GePb), calculated via DFT using the HSEsol (solid lines) and mBJ (dashed lines) XC functionals. All calculations include spin-orbit coupling. For comparative purposes, the zero of energy has been chosen to lie at the Fermi energy -- i.e.~the energy of the $\Gamma_{8v}$ VB edge states -- in all cases.}
	\label{fig:primitive_band_structures}
\end{figure*}

Table~\ref{tab:dft_benchmark} summarises the results of our DFT calculations for the constituent crystalline materials relevant to Ge$_{1-x}$Pb$_{x}$: diamond-structured semiconducting Ge ($x = 0$), diamond-structured semimetallic \cite{Hermann_PRB_2010} Pb (d-Pb; $x = 1$), and the fictitious semimetallic zinc blende IV-IV compound GePb (zb-GePb). Listed in Table~\ref{tab:dft_benchmark} are the calculated lattice constants $a$, direct band gaps $E_{g}$, and valence band (VB) spin-orbit splitting energies $\Delta_{\scalebox{0.6}{\textrm{SO}}}$, compared to (low temperature) experimental measurements and previous theoretical calculations. The band structures of Ge, d-Pb and zb-GePb, calculated using the HSEsol (solid lines) and mBJ (dashed lines) XC functionals, are shown respectively in Figs.~\ref{fig:primitive_band_structures}(a),~\ref{fig:primitive_band_structures}(b) and~\ref{fig:primitive_band_structures}(c). We note that zb-GePb -- equivalent to an ordered Ge$_{0.5}$Pb$_{0.5}$ ($x = 0.5$) alloy -- is semimetallic, with a large negative direct $\Gamma_{7c}$-$\Gamma_{8v}$ energy gap $-2.695$ eV ($-2.407$ eV) in the HSEsol (mBJ) calculation, suggesting that the band gap in Ge$_{1-x}$Pb$_{x}$ alloys can be expected both to close and become direct at some Pb composition(s) significantly below 50\%.


Since states originating from different wave vectors $\textbf{k}$ in the Brillouin zone of the primitive unit cell of the underlying diamond lattice are folded back to the zone centre (\textbf{K} = 0) in supercell calculations (see, e.g., Ref.~\onlinecite{Popescu_PRB_2012} and references therein), it can be difficult to identify the character of individual zone-centre states in the band structure of a Ge$_{1-x}$Pb$_{x}$ supercell, and hence to deduce the composition at which the alloy becomes a direct-gap semiconductor. To address this issue, we investigate how the alloy CB structure changes as a function of applied hydrostatic pressure. The pressure coefficients $\frac{ \textrm{d} E_{g} }{ \textrm{d} P }$ for the indirect L$_{6c}$-$\Gamma_{8v}$, direct $\Gamma_{7c}$-$\Gamma_{8v}$, and indirect X$_{5c}$-$\Gamma_{8v}$ band gaps of Ge are significantly different to one another having, e.g., respective values 4.66, 13.33 and $-$1.60 meV kbar$^{-1}$ in our HSEsol calculations. \cite{Halloran_OQE_2019,Eales_SR_2019} Calculation of $\frac{ \textrm{d} E_{g} }{ \textrm{d} P }$ for the fundamental band gap in a given alloy supercell then allows to identify the character of the band gap, \cite{Eales_SR_2019} and hence to track the evolution of the character of the CB edge states and band gap with increasing Pb composition $x$. We note that the quantitative information regarding band hybridisation contained in alloy supercell band gap pressure coefficients is implicitly contained in the spectral function underpinning zone unfolding approaches which are increasingly employed to analyse alloy electronic structure, \cite{Popescu_PRB_2012} but emphasise that direct calculation of pressure coefficients provides data which can be directly compared to experimental measurements. \cite{Halloran_OQE_2019,Eales_SR_2019}

Since, for small $x$, the Ge$_{1-x}$Pb$_{x}$ alloy CB edge states originate from the L$_{6c}$ CB edge states of Ge, we follow Ref.~\onlinecite{Halloran_OQE_2019} and restrict our attention to supercells in which hybridisation between L- and $\Gamma$-point eigenstates is permitted to occur. That is, we restrict our attention to $n \times n \times n$ face-centred cubic (FCC) or simple cubic (SC) supercells having even values of $n$. In such supercells the L points in the Brillouin zone of the underlying diamond primitive unit cell fold to the supercell zone centre $\textbf{K} = 0$, with L-point eigenstates then being free to hybridise with those at $\Gamma$ under the influence of the structural and chemical changes associated with Pb incorporation.


To investigate the electronic structure evolution in realistic, disordered Ge$_{1-x}$Pb$_{x}$ alloys we employ a series of 128-atom ($4 \times 4 \times 4$ FCC) SQSs, containing up to 10 Pb atoms -- i.e.~Pb compositions up to $x = 7.81$\%. The SQSs were generated stochastically using a Monte Carlo simulated annealing procedure -- as implemented in the Alloy Theoretic Automated Toolkit (ATAT) \cite{vandeWalle_JC_2002,vandeWalle_JC_2009,vandeWalle_JC_2013} -- to optimise the supercell lattice correlation functions up to third nearest-neighbour distance about each lattice site, with respect to the target lattice correlation functions for a randomly disordered, diamond-structured binary alloy having a specified Pb composition $x$. \cite{Zunger_PRL_1990,Wei_PRB_1990}


\section{Results}
\label{sec:results}

In this section we present the results of our theoretical analysis. We begin in Sec.~\ref{sec:results_ordered} with the impact of Pb incorporation on the electronic structure of ordered alloy supercells, and demonstrate the transition to a direct band gap with increasing Pb composition $x$. We compare the results of calculations undertaken using both the HSEsol and mBJ XC functionals, establishing the accuracy of the latter for use in further analysis. In Sec.~\ref{sec:results_clustering} we turn our attention to disordered supercells, and illustrate the importance of alloy disorder in Ge$_{1-x}$Pb$_{x}$ by tracking the evolution of the alloy CB edge as the separation between two Pb atoms reduces from fourth-nearest neighbour to forming a nearest-neighbour Pb-Pb pair. Finally, in Sec.~\ref{sec:results_disordered} we use 128-atom SQSs to analyse the evolution of the electronic structure with Pb composition $x$, and to quantify the nature of the indirect- to direct-gap transition in realistic, disordered Ge$_{1-x}$Pb$_{x}$ alloys.


\subsection{Conduction band structure and Pb-induced band mixing in ordered Ge$_{1-x}$Pb$_{x}$ supercells}
\label{sec:results_ordered}


We begin our analysis by considering the structural and electronic properties of ordered Ge$_{63}$Pb$_{1}$ ($x = 1.56$\%, $2 \times 2 \times 2$ SC) and Ge$_{15}$Pb$_{1}$ ($x = 6.25$\%, $2 \times 2 \times 2$ FCC) alloy supercells. Considering the data of Table~\ref{tab:dft_benchmark}, we note that the lattice constant of zb-GePb is lower than the average of those calculated for Ge and d-Pb, using both the LDA and HSEsol XC functionals. This suggests that the Ge$_{1-x}$Pb$_{x}$ alloy lattice constant deviates from that predicted based on V\'{e}gard's law (linear interpolation), and possesses a positive bowing coefficient $b$, where $a (x) = ( 1 - x ) \, a ( \textrm{Ge} ) + x \, a ( \textrm{d-Pb} ) - b \, x \, ( 1 - x )$ is the Ge$_{1-x}$Pb$_{x}$ alloy lattice constant. Examining the alloy lattice constants obtained via structural relaxation of the Ge$_{63}$Pb$_{1}$ supercell, we find that this is indeed the case. Using the LDA (HSEsol) XC functional we calculate $a = 5.666$ \AA~(5.667 \AA), which is lower than the value of 5.668 \AA~(5.673 \AA) obtained via linear interpolation between the calculated lattice constants of Ge and d-Pb. Using the LDA (HSEsol) relaxed lattice constants we then compute a bowing parameter $b = 0.13$ \AA~(0.39 \AA) for the Ge$_{1-x}$Pb$_{x}$ alloy. Similar trends are observed in the relaxed lattice constants for the Ge$_{15}$Pb$_{1}$ supercell. Overall, our calculations suggest bowing in the range $b \approx 0.1$ -- 0.4~\AA~for these ordered structures having low Pb compositions $x \leq 6.25$\%. We will see however in Sec.~\ref{sec:results_disordered} below, that this result is modified in the presence of alloy disorder, producing a change in the sign of $b$ computed for disordered SQS supercells.


\begin{table*}[t!]
	\caption{\label{tab:ordered_supercell_results} Fundamental band gap $E_{g}$, VB spin-orbit splitting energy $\Delta_{\protect\scalebox{0.6}{\textrm{SO}}}$, and band gap pressure coefficient $\frac{ \textrm{d} E_{g} }{ \textrm{d} P }$ for ordered Ge$_{63}$Pb$_{1}$ ($x = 1.56$\%) and Ge$_{15}$Pb$_{1}$ ($x = 6.25$\%) alloy supercells, calculated via DFT using the HSEsol and mBJ XC functionals. The corresponding calculated values of the direct $\Gamma_{7c}$-$\Gamma_{8v}$ and indirect (fundamental) L$_{6c}$-$\Gamma_{8v}$ band gaps of Ge are provided for reference, with the latter listed in parentheses. The results of equivalent calculations \cite{Halloran_OQE_2019} for ordered Ge$_{1-x}$Sn$_{x}$ supercells are provided for comparative purposes.}
	\begin{ruledtabular}
		\begin{tabular}{cc|cc|cc|cc}
			 & & $E_{g}$ (eV) & $E_{g}$ (eV) & $\Delta_{\scalebox{0.6}{\textrm{SO}}}$ (eV) & $\Delta_{\scalebox{0.6}{\textrm{SO}}}$ (eV) & $\frac{ \textrm{d} E_{g} }{ \textrm{d} P }$ (meV kbar$^{-1}$) & $\frac{ \textrm{d} E_{g} }{ \textrm{d} P }$ (meV kbar$^{-1}$) \\
			Supercell         & $x$ (\%) & HSEsol        & mBJ           & HSEsol      & mBJ         & HSEsol        & mBJ         \\
			\hline
			Ge                & -----    & 0.909 (0.766) & 0.868 (0.724) & 0.322       & 0.274       & 13.33 (4.66) & 13.23 (4.07) \\
			\hline
			Ge$_{63}$Pb$_{1}$ & 1.56     & 0.616         & 0.596         & 0.381       & 0.334       & 6.24         &  5.68        \\
			Ge$_{63}$Sn$_{1}$ & 1.56     & 0.681$^{a}$   & 0.660$^{a}$   & 0.334$^{a}$ & 0.282$^{a}$ & 4.75$^{a}$   &  4.19$^{a}$  \\
			\hline
			Ge$_{15}$Pb$_{1}$ & 6.25     & 0.020         & $-$0.040      & 0.612       & 0.571       & 10.23        &  9.17        \\
			Ge$_{15}$Sn$_{1}$ & 6.25     & 0.388$^{a}$   & 0.356$^{a}$   & 0.379$^{a}$ & 0.316$^{a}$ & 10.00$^{a}$  &  9.50$^{a}$
		\end{tabular}
	\end{ruledtabular}
	\begin{flushleft}
	$^{a}$Calc., Ref.~\onlinecite{Halloran_OQE_2019}
	\end{flushleft}
\end{table*}

\begin{figure*}[t!]
	\includegraphics[width=1.00\textwidth]{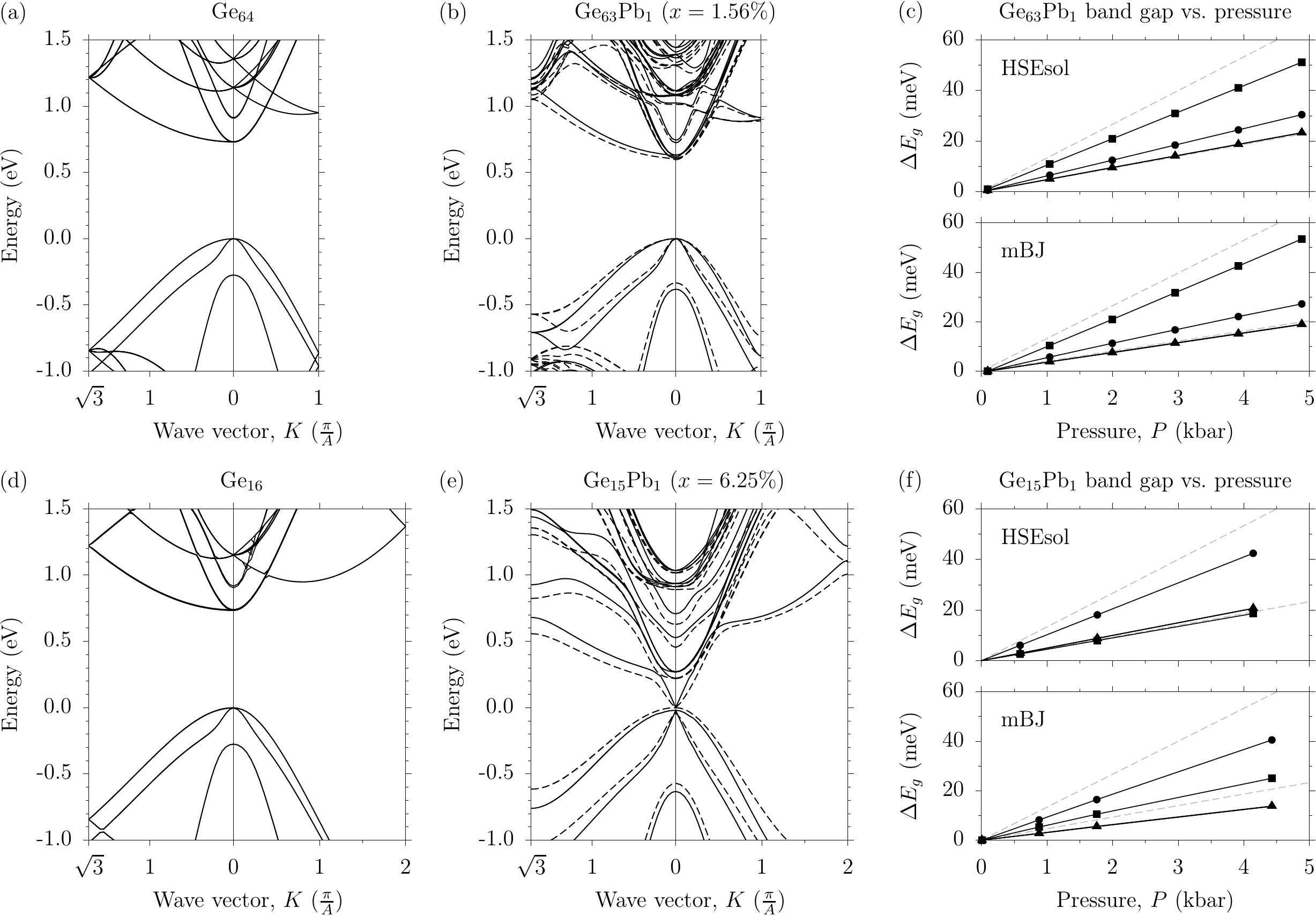}
	\caption{Band structure of (a) Ge$_{64}$ ($2 \times 2 \times 2$ SC), (b) Ge$_{63}$Pb$_{1}$ ($x = 1.56$\%), (d) Ge$_{16}$($2 \times 2 \times 2$ FCC), and (e) Ge$_{15}$Pb$_{1}$ ($x = 6.25$\%), calculated via DFT using the HSEsol (solid lines) and mBJ (dashed lines) XC functionals. (c) Calculated change in band gap $\Delta E_{g}$ with applied hydrostatic pressure $P$ for the Ge$_{63}$Pb$_{1}$ supercell of (b), calculated using the HSEsol (upper panel) and mBJ (lower panel) XC functionals. Band gaps are calculated between the VB edge and the first (circles), second, third and fourth (triangles), and fifth (squares) lowest energy CB states. (f) As in (c), but for the Ge$_{15}$Pb$_{1}$ supercell of (e). The lower (upper) dashed grey line in (c) and (f) denotes the variation with pressure of the indirect L$_{6c}$-$\Gamma_{8v}$ (direct $\Gamma_{7c}$-$\Gamma_{8v}$) band gap of Ge.}
	\label{fig:ordered_supercell_band_structures}
\end{figure*}

Turning our attention now to the electronic structure of these ordered supercells, Table~\ref{tab:ordered_supercell_results} shows the calculated fundamental band gaps $E_{g}$, VB spin-orbit splitting energies $\Delta_{\scalebox{0.6}{\textrm{SO}}}$, and fundamental band gap pressure coefficients $\frac{ \textrm{d} E_{g} }{ \textrm{d} P }$. The results of equivalent calculations \cite{Halloran_OQE_2019} for Ge$_{1-x}$Sn$_{x}$ supercells are provided for comparative purposes. The HSEsol- (mBJ-) calculated band structures of the Ge$_{63}$Pb$_{1}$ and Ge$_{15}$Pb$_{1}$ supercells are shown using solid (dashed) lines in Figs.~\ref{fig:ordered_supercell_band_structures}(b) and~\ref{fig:ordered_supercell_band_structures}(e), respectively. For reference, the band structures of the corresponding Pb-free Ge$_{64}$ and Ge$_{16}$ supercells are shown in Figs.~\ref{fig:ordered_supercell_band_structures}(a) and~\ref{fig:ordered_supercell_band_structures}(d). Since the 64-atom supercells have SC lattice vectors, we note that the zone boundary along (001) lies at supercell wave vector $K_{z} = \frac{ \pi }{ A }$, where $A = 2a$ is the supercell lattice constant for a 64-atom $2 \times 2 \times 2$ SC supercell. The corresponding zone boundary lies at $K_{z} = \frac{ 2 \pi }{ A }$ for a 16-atom $2 \times 2 \times 2$ FCC supercell.

For the Ge$_{63}$Pb$_{1}$ supercell the HSEsol- (mBJ-) calculated band gap $E_{g} = 0.616$ eV (0.596 eV) is reduced by 150 meV (128 meV) compared to the fundamental band gap of Ge. We also calculate a strong increase in $\Delta_{\scalebox{0.6}{\textrm{SO}}}$ of 59 meV (60 meV) due to Pb incorporation, with $\Delta_{\scalebox{0.6}{\textrm{SO}}}$ increasing from 0.322 eV (0.274 eV) in Ge, to 0.381 eV (0.334 eV) in Ge$_{63}$Pb$_{1}$. These results equate to a very strong decrease of $E_{g}$, of $\approx 80$ -- 100 meV per \% Pb, and a strong increase in $\Delta_{\scalebox{0.6}{\textrm{SO}}}$, of $\approx 40$ meV per \% Pb. Indeed, for the Ge$_{15}$Pb$_{1}$ supercell the HSEsol- (mBJ-) calculated band gap $E_{g} = 0.020$ eV ($-0.040$ eV) indicates a closing of the alloy band gap for Pb compositions as low as $x = 6.25$\%. The corresponding VB spin-orbit splitting energies are $\Delta_{\scalebox{0.6}{\textrm{SO}}} = 0.612$ eV (0.571 eV). We note that these values are significantly in excess of the changes in $E_{g}$ and $\Delta_{\scalebox{0.6}{\textrm{SO}}}$ associated with Sn incorporation (cf.~Table~\ref{tab:ordered_supercell_results}), reflecting the larger differences in size and chemical properties between Ge and Pb than between Ge and Sn. While the mBJ XC functional tends to give slightly reduced $E_{g}$ and $\Delta_{\scalebox{0.6}{\textrm{SO}}}$ compared to the HSEsol XC functional, we note that the changes in $E_{g}$ and $\Delta_{\scalebox{0.6}{\textrm{SO}}}$ in response to Pb incorporation calculated using both approaches are in good quantitative agreement.


Considering now the calculated supercell band dispersion in Figs.~\ref{fig:ordered_supercell_band_structures}(b) and~\ref{fig:ordered_supercell_band_structures}(e) -- shown respectively using solid and dashed lines for calculations employing the HSEsol and mBJ XC functionals -- we note that Pb incorporation has a significant impact on the CB structure, while the primary impact close in energy to the VB edge is, as noted above, to increase the spin-orbit splitting energy $\Delta_{\scalebox{0.6}{\textrm{SO}}}$. We find that the evolution of the band structure close in energy to the VB edge can be well described using a conventional alloy approach (e.g. the VCA, with appropriate bowing coefficients for the VB edge energy and $\Delta_{\scalebox{0.6}{\textrm{SO}}}$). This, as we will describe in detail below, is not the case for the CB structure.


In Ge$_{64}$ and Ge$_{16}$ the CB states at the supercell zone centre $\textbf{K} = 0$ are, in order of increasing energy: (i) the eightfold (fourfold and Kramers) degenerate folded L$_{6c}$ L-point CB minimum states, (ii) the twofold (Kramers) degenerate $\Gamma_{7c}$ zone-centre CB edge states, and (iii) the twelvefold (sixfold and Kramers) degenerate folded X$_{5c}$ X-point CB edge states. In Ge$_{63}$Pb$_{1}$ we note a downward shift in energy, and lifting of the degeneracy, of the alloy CB edge states originating from the folded L$_{6c}$ states of the Ge host matrix. We also calculate a significant reduction in energy of the alloy CB states originating from the Ge $\Gamma_{7c}$ host matrix states, which is larger in magnitude than that associated with the alloy CB states originating from the L$_{6c}$ states of Ge. We find that the lowest energy CB states in Ge$_{63}$Pb$_{1}$ comprise a Kramers degenerate singlet, possessing purely $s$-like orbital character ($A_{1}$ symmetry) at the Pb lattice site. The second lowest energy CB states in this supercell lie 14 meV (11 meV) higher in energy in the HSEsol (mBJ) calculation, and constitute a sixfold (threefold and Kramers) degenerate triplet possessing purely $p$-like orbital character ($T_{2}$ symmetry) at the Pb lattice site. The substitutional incorporation of the Pb atom also allows hybridisation between different (linear combinations of) Ge host matrix states of the same symmetry, \cite{Lindsay_PB_2003} so that the lowest energy CB state in Ge$_{63}$Pb$_{1}$ can be described in terms of a linear combination of Ge host matrix states which lie close in energy to the CB edge and possess $A_{1}$ symmetry at the Pb lattice site. \cite{Broderick_NUSOD_2019} We note that the splitting of the alloy CB states originating from the L$_{6c}$ states of Ge -- into a singlet having $s$-like symmetry at the impurity site, and a triplet having $p$-like symmetry at the impurity site -- is qualitatively similar to that in a Ge$_{63}$Sn$_{1}$ supercell. \cite{Halloran_OQE_2019,Broderick_GeSn_preparation_2019} However, the ordering of these $A_{1}$- and $T_{2}$-symmetric Ge L$_{6c}$-derived alloy CB edge states is reversed in Ge$_{63}$Pb$_{1}$ compared to Ge$_{63}$Sn$_{1}$. In Ge$_{15}$Pb$_{1}$ we note the same trends as in Ge$_{63}$Pb$_{1}$ in terms of the symmetry, degeneracy and ordering of the lowest energy alloy CB states: the (Kramers degenerate) Ge$_{15}$Pb$_{1}$ alloy CB edge is again a singlet possessing purely $s$-like orbital character at the Pb lattice site, while the second lowest energy CB states constitute a triplet possessing purely $p$-like orbital character at the Pb lattice site. We note however a significant increase of the separation in energy -- 269 meV (260 meV) in the HSEsol (mBJ) calculation -- between these $A_{1}$-symmetric singlet and $T_{2}$-symmetric triplet states.


Consideration of the alloy band structure therefore indicates the presence of Pb-induced and Pb composition-dependent band mixing (hybridisation) in the CB of the ordered Ge$_{1-x}$Pb$_{x}$ supercells investigated, qualitatively similar to that observed in Ge$_{1-x}$Sn$_{x}$. \cite{Halloran_OQE_2019} To further investigate the presence and extent of Pb-induced band mixing in these supercells, we have calculated the pressure coefficients associated with the band gaps between the alloy VB edge and the five lowest energy alloy CB states (which originate predominantly from the L$_{6c}$ and $\Gamma_{7c}$ states of Ge). Since we are concerned with low Pb compositions $x < 10$\%, it is expected that an alloy having an indirect (direct) band gap will have a pressure coefficient close to that of the corresponding indirect L$_{6c}$-$\Gamma_{8v}$ (direct $\Gamma_{7c}$-$\Gamma_{8v}$) band gap of Ge. Before considering the calculated values of $\frac{ \textrm{d} E_{g} }{ \textrm{d} P }$ for the alloy supercells of Fig.~\ref{fig:ordered_supercell_band_structures}, we note that our respective HSEsol- (mBJ-) calculated pressure coefficients $\frac{ d E_{g} }{ d P } = 4.66$ meV kbar$^{-1}$ (4.07 meV kbar$^{-1}$) and 13.33 meV kbar$^{-1}$ (13.23 meV kbar$^{-1}$) for the indirect and direct band gaps of Ge (cf.~Table~\ref{tab:ordered_supercell_results}) are in good quantitative agreement with the measured values of 4.3 and 12.9 meV kbar$^{-1}$. \cite{Eales_SR_2019}

To compute $\frac{ \textrm{d} E_{g} }{ \textrm{d} P }$ we apply hydrostatic pressure by compressing the lattice vectors of a given relaxed alloy supercell, and then perform geometric (re-)optimisation by allowing only the internal degrees of freedom (ionic positions) to relax. The results of our calculations of $\frac{ \textrm{d} E_{g} }{ \textrm{d} P }$ are summarised in Figs.~\ref{fig:ordered_supercell_band_structures}(c) and~\ref{fig:ordered_supercell_band_structures}(f), for the Ge$_{63}$Pb$_{1}$ and Ge$_{15}$Pb$_{1}$ supercells respectively. In each case the upper (lower) panel shows the results obtained using the HSEsol (mBJ) XC functional; closed circles, triangles and squares respectively show the pressure coefficients associated with the band gaps calculated between the VB edge and the first, second to fourth, and fifth lowest energy CB states. The lower and upper dashed grey lines in Figs.~\ref{fig:ordered_supercell_band_structures}(c) and~\ref{fig:ordered_supercell_band_structures}(f) respectively show the corresponding calculated pressure dependence of the Ge L$_{6c}$-$\Gamma_{8v}$ and $\Gamma_{7c}$-$\Gamma_{8v}$ band gaps.

For both supercells we observe that the pressure coefficient associated with the band gap calculated using the second to fourth lowest energy alloy CB states (closed red triangles) -- i.e.~the triplet states which are $T_{2}$-symmetric about the Pb lattice site -- remains approximately equal to that of the Ge L$_{6c}$-$\Gamma_{8v}$ indirect fundamental band gap. This confirms that these triplet states undergo minimal hybridisation with other Ge states, and are comprised almost entirely of a linear combination of Ge L$_{6c}$ states possessing $p$-like orbital character at the Pb lattice site. Conversely, for Ge$_{63}$Pb$_{1}$ the HSEsol- (mBJ-) calculated fundamental supercell band gap pressure coefficient $\frac{ \textrm{d} E_{g} }{ \textrm{d} P } = 6.24$ meV kbar$^{-1}$ (5.68 meV kbar$^{-1}$), denoted using closed circles in Figs.~\ref{fig:ordered_supercell_band_structures}(c) and~\ref{fig:ordered_supercell_band_structures}(f), is increased by $\approx 1.6$ meV kbar$^{-1}$ compared to that of the Ge L$_{6c}$-$\Gamma_{8v}$ fundamental indirect band gap. This, combined with the symmetry of the alloy CB states, suggests the presence of Pb-induced hybridisation in the supercell between a singlet state formed of an $A_{1}$-symmetric linear combination of Ge L$_{6c}$ states (possessing purely $s$-like orbital character at the Pb lattice site), and the Ge $\Gamma_{7c}$ zone-centre CB edge state (which possesses $s$-like orbital character at all lattice sites).

The presence of this Pb-induced hybridisation is confirmed by noting that $\frac{ \textrm{d} E_{g} }{ \textrm{d} P }$ associated with the band gap between the VB edge and the fifth highest energy CB state -- which originates from the Ge $\Gamma_{7c}$ states, denoted using closed squares in Figs.~\ref{fig:ordered_supercell_band_structures}(c) and~\ref{fig:ordered_supercell_band_structures}(f) -- is reduced compared to that of the Ge $\Gamma_{7c}$-$\Gamma_{8v}$ direct band gap, suggesting the acquisition of an admixture of Ge L$_{6c}$ (and possibly also a minor admixture of Ge X$_{5c}$) character. Considering the Ge$_{15}$Pb$_{1}$ pressure coefficients in Fig.~\ref{fig:ordered_supercell_band_structures}(f), we see that the pressure coefficients associated with the band gaps between the VB edge and the second to fourth lowest energy ($T_{2}$-symmetric triplet) CB states are again very close to that associated with the Ge L$_{6c}$-$\Gamma_{8v}$, but that the largest pressure coefficient of 10.23 meV kbar$^{-1}$ (9.17 meV kbar$^{-1}$) is now associated with the CB edge state (closed circles), indicating the emergence of a fundamental alloy band gap which is hybridised but primarily direct in nature (with some Ge $\Gamma_{7c}$ character continuing to reside on the fifth lowest energy CB states at $x = 6.25$\%).


The observed $\Gamma_{7c}$-L$_{6c}$ hybridisation occurs in part due to the small $\approx 145$ meV separation in energy between the L$_{6c}$ and $\Gamma_{7c}$ states in Ge, and is qualitatively similar to that observed in equivalent analysis of Ge$_{1-x}$Sn$_{x}$ alloy supercells. \cite{Halloran_OQE_2019} Overall, our analysis of idealised (ordered) alloy supercells demonstrates that the electronic structure evolution and indirect- to direct-gap transition in Ge$_{1-x}$Pb$_{x}$ is qualitatively similar to that in Ge$_{1-x}$Sn$_{x}$, but with alloy band mixing effects being less pronounced in Ge$_{1-x}$Pb$_{x}$ compared to equivalent Ge$_{1-x}$Sn$_{x}$ supercells. Indeed, as we will demonstrate in Sec.~\ref{sec:results_disordered} below, when alloy disorder is taken into account alloy band mixing effects associated with Pb incorporation are significantly less pronounced than the corresponding effects in equivalent Ge$_{1-x}$Sn$_{x}$ supercells, and hence play a less important role in determining the nature and evolution of the alloy electronic structure with increasing $x$. Finally, having established here that the mBJ XC functional gives results in good quantitative agreement with the more computationally expensive HSEsol XC functional, we employ the former for the remainder of our analysis.


\begin{figure*}[t!]
	\includegraphics[width=1.00\textwidth]{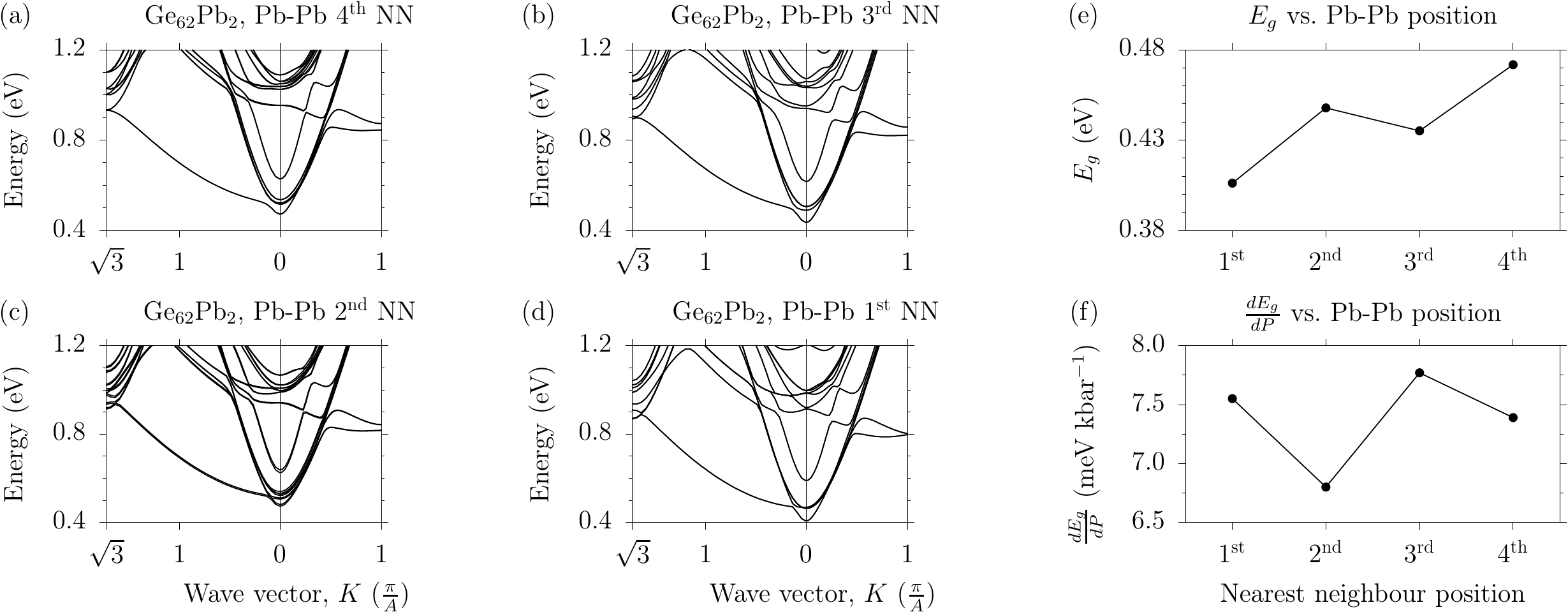}
	\caption{Calculated CB structure of disordered, 64-atom ($2 \times 2 \times 2$ SC) Ge$_{62}$Pb$_{2}$ ($x = 3.125$\%) supercells in which the two Pb atoms are (a) fourth, (b) third, (c) second, and (d) first nearest neighbours (NNs). (e) Calculated variation of the fundamental band gap $E_{g}$ of the Ge$_{62}$Pb$_{2}$ supercells of (a) -- (d) with respect to the relative position of the two Pb atoms. (f) as in (e), but for the pressure coefficient $\frac{ \textrm{d} E_{g} }{ \textrm{d} P }$ associated with the fundamental band gap $E_{g}$.}
	\label{fig:disordered_supercell_band_structures}
\end{figure*}


\subsection{Impact of Pb local environment on Ge$_{1-x}$Pb$_{x}$ alloy electronic structure}
\label{sec:results_clustering}


Having quantified the impact of Pb incorporation on the band structure of ordered Ge$_{1-x}$Pb$_{x}$ supercells, we turn our attention now to the impact of short-range disorder (Pb clustering) on the electronic structure. We begin with a Ge$_{63}$Pb$_{1}$ supercell and substitute a second Pb atom to form four distinct Ge$_{62}$Pb$_{2}$ ($x = 3.125$\%) supercells in which the two Pb atoms are located at fourth-, third-, second- and first-nearest neighbour lattice sites. In each of these four cases we calculate the electronic structure and quantify the dependence of the fundamental band gap $E_{g}$, and its pressure coefficient $\frac{ \textrm{d} E_{g} }{ \textrm{d} P }$, on the relative position of the two Pb atoms in the supercell. We note that incorporation of $> 1$ substitutional Pb atoms breaks the cubic symmetry of the underlying diamond lattice, which was preserved in the ordered supercell calculations of Sec.~\ref{sec:results_ordered}. As a result of this reduction in symmetry the fourfold (twofold and Kramers) degeneracy of the zone-centre VB edge states is lifted, giving rise to two distinct sets of Kramers degenerate states (which, in general, possess an admixture of light- and heavy-hole-like Ge $\Gamma_{8v}$ character). The precise magnitude of this VB edge splitting in a disordered alloy supercell is in general a non-monotonic function of alloy composition, but instead depends on the precise short-range disorder present in a given supercell. In this section, as well as in Sec.~\ref{sec:results_disordered} below, we therefore calculate energy gaps in disordered supercells with respect to the average energy of the split VB edge states.


The results of our calculations -- using the mBJ XC functional -- for this series of Ge$_{62}$Pb$_{2}$ supercells are summarised in Fig.~\ref{fig:disordered_supercell_band_structures}, where Figs.~\ref{fig:disordered_supercell_band_structures}(a),~\ref{fig:disordered_supercell_band_structures}(b),~\ref{fig:disordered_supercell_band_structures}(c) and~\ref{fig:disordered_supercell_band_structures}(d) respectively show the calculated Ge$_{62}$Pb$_{2}$ CB structure in the case of having fourth-, third-, second- and first-nearest neighbour Pb atoms. Figures~\ref{fig:disordered_supercell_band_structures}(e) and~\ref{fig:disordered_supercell_band_structures}(f) respectively show the dependence of the calculated values of $E_{g}$ and $\frac{ \textrm{d} E_{g} }{ \textrm{d} P }$ on the relative position of the two Pb atoms. The band structures shown in Figs.~\ref{fig:disordered_supercell_band_structures}(a) --~\ref{fig:disordered_supercell_band_structures}(d) display, overall, strong qualitative similarity. As in the ordered supercells described above, the (Kramers degenerate) CB edge in each supercell constitutes a singlet possessing $s$-like orbital character at Pb lattice sites, while the next three (Kramers degenerate) CB states constitute a triplet possessing $p$-like orbital character at Pb lattice sites. While in an ordered alloy supercell these $p$-like Ge L$_{6c}$-derived states form a degenerate triplet, their degeneracy is lifted in the supercells considered here due to loss of the underlying cubic symmetry of the lattice. However, as we describe below, despite close qualitative similarities in the calculated band structures, Pb clustering produces pronounced quantitative differences in the calculated electronic properties in supercells having fixed Pb composition $x$.


Beginning with a Ge$_{62}$Pb$_{2}$ supercell in which the two Pb atoms are fourth-nearest neighbours we calculate a fundamental band gap $E_{g} = 0.472$ eV, which is reduced by 252 meV compared to the fundamental band gap of Ge, and by 124 meV compared that of the ordered Ge$_{63}$Pb$_{1}$ supercell considered in Sec.~\ref{sec:results_ordered}. Correspondingly, we calculate $\Delta_{\scalebox{0.6}{\textrm{SO}}} = 0.377$ eV for this Ge$_{62}$Pb$_{2}$ supercell, which is increased by 103 meV compared to that in Ge, and by 43 meV compared to that in Ge$_{63}$Pb$_{1}$. In line with the ordered supercell calculations of Sec.~\ref{sec:results_ordered}, these results confirm that Pb incorporation in Ge leads to a strong decrease of $E_{g}$ and increase of $\Delta_{\scalebox{0.6}{\textrm{SO}}}$ with increasing $x$. For the case of third-nearest neighbour Pb atoms we calculate $E_{g} = 0.435$ eV ($\Delta_{\scalebox{0.6}{\textrm{SO}}} = 0.396$ eV), which is 37 meV lower (19 meV higher) than that calculated in the case of fourth-nearest neighbour Pb atoms. Ultimately, in the presence of a nearest-neighbour Pb-Pb pair we calculate $E_{g} = 0.406$ eV, which is reduced by 318 meV compared to the fundamental band gap of Ge, and by 66 meV compared to the case of fourth-nearest neighbour Pb atoms. Similarly, we calculate $\Delta_{\scalebox{0.6}{\textrm{SO}}} = 0.425$ eV for the supercell containing a Pb-Pb nearest-neighbour pair, which is increased by 151 meV compared to that in Ge, and by 48 meV compared to the case of fourth-nearest neighbour Pb atoms. These calculations indicate that the formation of Pb clusters, in the form of substitutional Pb atoms occupying near-neighbour lattice sites, tends to strongly decrease $E_{g}$ and increase $\Delta_{\scalebox{0.7}{\textrm{SO}}}$. We note that the calculated decrease of $E_{g}$ (cf.~Fig.~\ref{fig:disordered_supercell_band_structures}(e)) and increase of $\Delta_{\scalebox{0.6}{\textrm{SO}}}$ (not shown) as the two Pb atoms are substituted on successively closer lattice sites is not a monotonic function of the Pb-Pb interatomic distance, but depends on the specific relative position of the two Pb atoms. Since these four supercells have equal Pb composition $x = 3.125$\%, this calculated $\approx 70$ meV variation in $E_{g}$ -- representing $\approx 15$\% of the total band gap -- highlights the significant impact that Pb-Pb clustering can have on the electronic properties of Ge$_{1-x}$Pb$_{x}$ alloys.


Figure~\ref{fig:disordered_supercell_band_structures}(f) shows the calculated pressure coefficients $\frac{ \textrm{d} E_{g} }{ \textrm{d} P }$, associated with the fundamental supercell band gap $E_{g}$, as a function of Pb-Pb separation for the same series of Ge$_{62}$Pb$_{2}$ supercells. In line with the ordered supercell calculations of Sec.~\ref{sec:results_ordered}, the calculated values of $\frac{ \textrm{d} E_{g} }{ \textrm{d} P }$ are intermediate between the values 4.07 and 13.23 meV kbar$^{-1}$ associated with the L$_{6c}$-$\Gamma_{8v}$ and $\Gamma_{7c}$-$\Gamma_{8v}$ band gaps of Ge (cf.~Table~\ref{tab:ordered_supercell_results}). In all four cases -- having fourth-, third-, second- and first-nearest neighbour Pb atoms -- the calculated value of $\frac{ \textrm{d} E_{g} }{ \textrm{d} P }$ remains closer to that associated with the Ge fundamental (indirect) band gap, suggesting that the supercell band gap remains primarily indirect in nature, but with some direct ($\sim 30$ -- 40\% Ge $\Gamma_{7c}$) character, at $x = 3.125$\%.  This observation of a hybridised alloy band gap is to be compared with previous theoretical analysis of Ge$_{1-x}$Pb$_{x}$ alloys, which have alternatively predicted a band gap having purely indirect \cite{Huang_PB_2014} or direct \cite{Huang_JAC_2017} character at this composition. Although our calculations show a Kramers degenerate singlet at the CB edge (cf.~Figs.~\ref{fig:disordered_supercell_band_structures}(a) --~\ref{fig:disordered_supercell_band_structures}(d)), the band gap pressure coefficient shows in each case that the band gap retains primarily indirect character. We note that the value of $\frac{ \textrm{d} E_{g} }{ \textrm{d} P }$ also varies non-monotonically with Pb-Pb separation, revealing that the details of the Pb-induced alloy band mixing are impacted by the specific short-range disorder present in a given alloy supercell. This demonstrates more generally that the character of the CB edge states, and hence the nature of the band gap, in real (disordered) Ge$_{1-x}$Pb$_{x}$ alloys, is sensitive to the presence of short-range alloy disorder and Pb clustering.


Overall, these results demonstrate that the calculated values of $E_{g}$ and $\frac{ \textrm{d} E_{g} }{ \textrm{d} P }$ -- the latter reflecting the hybridised character of the alloy CB edge states -- as well as $\Delta_{\scalebox{0.7}{\textrm{SO}}}$, in Ge$_{1-x}$Pb$_{x}$ at fixed Pb composition $x$ display significant dependence on the precise short-range disorder present in a given alloy supercell. We note however that due to the small number of atoms in these supercells, the importance of such short-range disorder effects may be somewhat exaggerated here in comparison to the effects that would likely be noted in larger scale alloy supercell calculations. This issue is discussed in more detail in Sec.~\ref{sec:results_disordered} below. While our analysis in Sec.~\ref{sec:results_ordered} identified the presence of Pb-induced band mixing in small Ge$_{1-x}$Pb$_{x}$ supercells, our calculations here indicate the interplay of these effects with short-range alloy disorder (particularly in the form of clustering of substitutional Pb atoms). From a theoretical perspective, our results therefore emphasise the breakdown of the VCA in Ge$_{1-x}$Pb$_{x}$ alloys, which neglects effects related to alloy band mixing and (short- or long-range) alloy disorder. We therefore conclude that atomistic calculations, which explicitly account for the differences in size and chemical properties between Ge and Pb, are required to provide quantitative insight into the properties of real Ge$_{1-x}$Pb$_{x}$ alloys.


\subsection{Disordered alloys: electronic structure evolution in Ge$_{1-x}$Pb$_{x}$ special quasi-random structures}
\label{sec:results_disordered}

\begin{figure*}[t!]
	\includegraphics[width=1.00\textwidth]{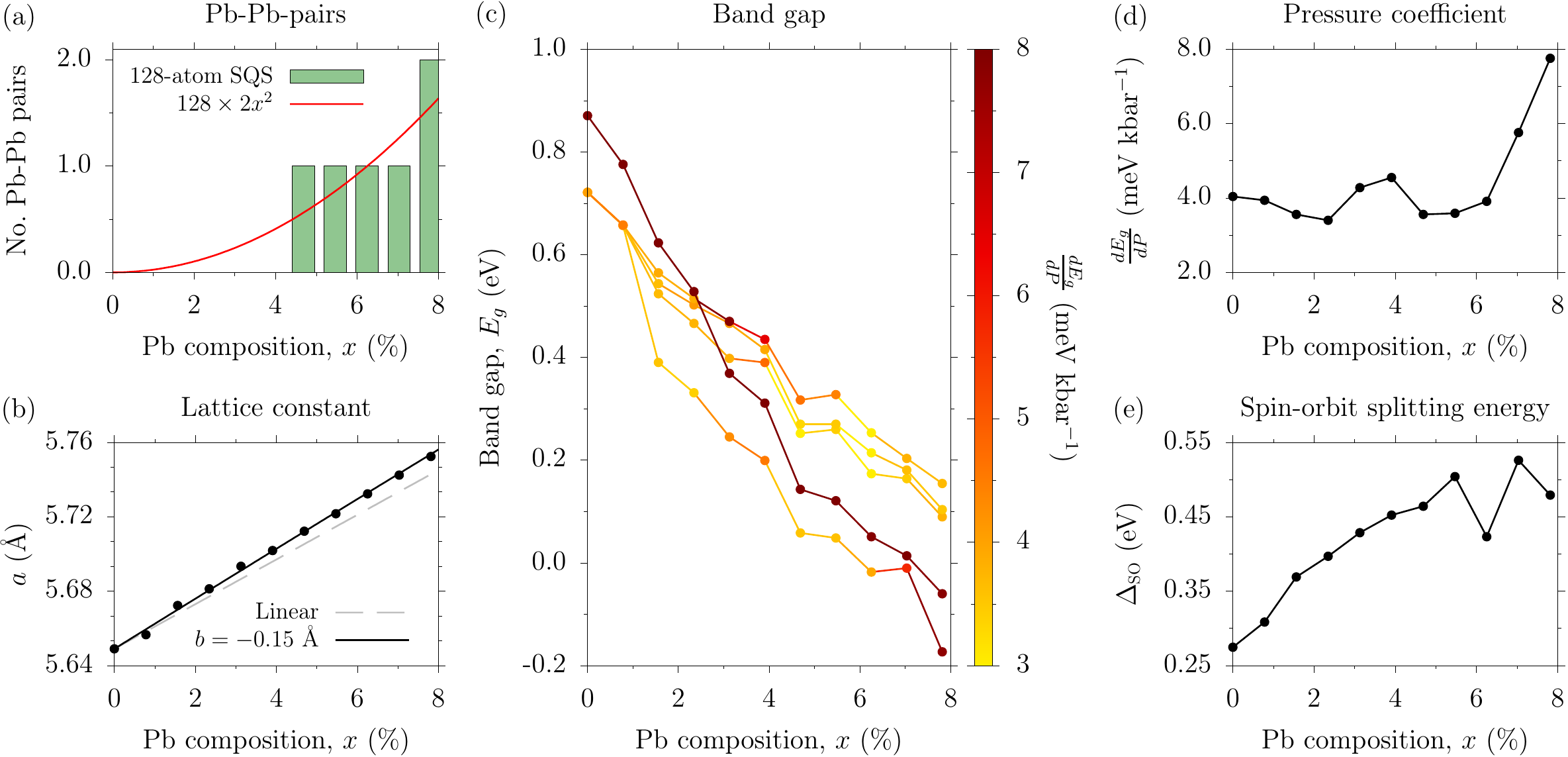}
	\caption{(a) Expected (solid red line) and actual (solid green bars) occurrence of nearest-neighbour Pb-Pb pairs in the 128-atom ($4 \times 4 \times 4$ FCC) SQSs used to analyse the electronic structure evolution in disordered Ge$_{1-x}$Pb$_{x}$ alloys. (b) Variation of the relaxed lattice constant $a$ with Pb composition $x$ (closed circles) for the SQSs of (a). The dashed grey line shows $a$ calculated by interpolating linearly between the lattice constants of Ge and diamond-structured Pb (cf.~Table~\ref{tab:dft_benchmark}). The solid black line shows the best fit to the calculated lattice constants, having bowing parameter $b = -0.15$ \AA. (c) Variation with $x$ of the band gaps $E_{g}$, calculated between the VB edge and the five lowest energy supercell zone centre CB states (closed circles). The colour of the data points are weighted according to the pressure coefficient $\frac{ \textrm{d} E_{g} }{ \textrm{d} P }$ associated with the corresponding band gap $E_{g}$. (d) Variation of $\frac{ \textrm{d} E_{g} }{ \textrm{d} P }$ with $x$ for the fundamental band gap of (c). (d) Variation of the VB spin-orbit splitting energy $\Delta_{\protect\scalebox{0.6}{\textrm{SO}}}$ with $x$.}
	\label{fig:sqs_summary}
\end{figure*}

Having established the importance of short-range Pb-related structural disorder in determining the details of the alloy electronic structure, we turn our attention now to the evolution of the electronic structure with $x$ in disordered Ge$_{1-x}$Pb$_{x}$ SQSs. We begin by assessing the suitability of the SQSs used in our electronic structure calculations, in terms of the occurrence of clusters of neighbouring Pb atoms. In a randomly disordered substitutional Ge$_{1-x}$Pb$_{x}$ alloy having Pb composition $x$, the probability of occurrence of a nearest-neighbour Pb-Pb pair -- i.e.~the probability that two Pb atoms substituted at randomly chosen lattice sites occupy nearest-neighbour lattice sites -- is, for small $x$, $\approx$ $2x^{2}$. As such, we would expect that a randomly disordered $N$-atom supercell contains on a average $\approx N \times 2x^{2}$ Pb-Pb pairs. Figure~\ref{fig:sqs_summary}(a) shows the expected (solid red line) and actual (solid green bars) number of Pb-Pb pairs in the 128-atom SQSs used in our electronic structure calculations. The actual number of Pb-Pb pairs that occur in the SQSs is in all cases found to equal the expected number $128 \times 2x^{2}$, rounded to the nearest whole number. We note that the probability of occurrence of a larger cluster containing three nearest-neighbour Pb atoms is, in the composition range of interest, $\propto x^{3}$, which remains $\lesssim 10^{-3}$ (so that $N \times x^{3} \sim 10^{-1}$ here) in the composition range considered. Hence we expect, and find, that no Pb clusters containing more than two neighbouring Pb atoms occur in the SQSs employed in our analysis. On this basis, we conclude that the 128-atom SQSs considered here have appropriate distributions of Pb atoms to analyse the evolution of the properties of randomly disordered Ge$_{1-x}$Pb$_{x}$ alloys with $x$.


The closed circles in Fig.~\ref{fig:sqs_summary}(b) show the variation of the (LDA-)relaxed alloy lattice constant $a$ with $x$ for the series of SQSs investigated. Comparing the calculated values of $a$ to the increase expected based on V\'{e}gard's law (dashed grey line), we observe that the lattice constants calculated for SQSs containing $> 1$ Pb atom exceed those expected based on a linear interpolation of the Ge and d-Pb lattice constants. This suggests a negative bowing coefficient $b$ for the lattice constant of disordered Ge$_{1-x}$Pb$_{x}$. Fitting to the calculated lattice constants -- depicted by the solid black line in Fig.~\ref{fig:sqs_summary}(b) -- yields a best-fit bowing coefficient $b = -0.15$ \AA. We note that this is in contrast to the results of our calculations in Sec.~\ref{sec:results_ordered}, where relaxation of ordered alloy supercells predicted a positive value of $b$ in the range 0.1 -- 0.4 \AA, reflecting that the presence of Pb-Pb pairs and disorder leads to larger local expansion of the crystal lattice (giving rise to a more rapid increase in $a$ with increasing $x$ in the presence of short-range disorder). Given that (i) real Ge$_{1-x}$Pb$_{x}$ alloy samples will inevitably contain varying degrees of short-range disorder, and (ii) SQSs are generally found to accurately predict the evolution with composition of the structural and elastic properties of randomly disordered alloys, \cite{vonPezold_PRB_2010} we expect that $b = -0.15$ \AA~represents a realistic estimate of the lattice constant bowing in Ge$_{1-x}$Pb$_{x}$ alloys. We note that this bowing parameter is close to that calculated for Ge$_{1-x}$Sn$_{x}$ alloys by Eckhardt et al. \cite{Eckhardt_PRB_2014}


Turning our attention to the electronic structure evolution, the closed circles in Fig.~\ref{fig:sqs_summary}(c) shows the calculated variation with $x$ of the band gaps between the (averaged) alloy VB edge and the five lowest energy alloy CB states. Here, we show the band gaps calculated for the five (Kramers degenerate) lowest energy CB states in each SQS: for $x = 0$ the four lowest energy states are the folded L$_{6c}$ CB minimum states of Ge, while the fifth is the $\Gamma_{7c}$ zone-centre CB edge state of Ge. The calculated reduction in fundamental band gap for this set of SQSs is $\approx 100$ meV per \% Pb replacing Ge, confirming that Pb incorporation drives extremely strong band gap reduction (comparable in magnitude to that observed in highly-mismatched III-V semiconductor alloys). At first glance, Fig.~\ref{fig:sqs_summary}(c) seems to suggest that an indirect- to direct-gap transition occurs for $x \approx 2\%$, around which composition we observe four (Kramers degenerate) CB states lying within $\approx 100$ meV of one another, with one (Kramers degenerate) state lying $\approx 150$ meV below these four higher energy states. However, we have weighted the colour of each of the data points in Fig.~\ref{fig:sqs_summary}(c) according to the corresponding calculated band gap pressure coefficient $\frac{ \textrm{d} E_{g} }{ \textrm{d} P }$: starting from yellow for $\frac{ \textrm{d} E_{g} }{ \textrm{d} P } = 3$ meV kbar$^{-1}$ (close to that of the Ge L$_{6c}$-$\Gamma_{8v}$ fundamental band gap), and shifting towards red as $\frac{ \textrm{d} E_{g} }{ \textrm{d} P }$ increases -- i.e.~as the direct (Ge $\Gamma_{7c}$) character of the alloy CB edge increases. Considering $\frac{ \textrm{d} E_{g} }{ \textrm{d} P }$ associated with each of the calculated band gaps, for $x \approx 2$\% we find that the four lowest energy band gaps have low associated pressure coefficients $\frac{ \textrm{d} E_{g} }{ \textrm{d} P } \approx 4$ meV kbar$^{-1}$, confirming that the four lowest energy alloy CB states at that composition retain predominantly indirect (Ge L$_{6c}$) character. At this composition the band gap associated with the fifth lowest energy CB state retains a large pressure coefficient $\frac{ \textrm{d} E_{g} }{ \textrm{d} P } \approx 10$ meV kbar$^{-1}$, confirming that this state retains primarily direct (Ge $\Gamma_{7c}$) character. As such, the alloy CB edge, and hence the alloy band gap, remains primarily indirect in nature for $x \approx 2$\%.

The general trends observed in Fig.~\ref{fig:sqs_summary}(c) are largely consistent with those observed for ordered supercells in Sec.~\ref{sec:results_ordered}, where we found that the CB states originating from the L$_{6c}$ states of Ge split into (Kramers degenerate) singlet and triplet states, respectively possessing $s$- and $p$-like orbital character at Pb lattice sites, with the singlet states lying lower in energy.  Due to the loss of cubic symmetry in the disordered supercells considered here we observe lifting of the degeneracy of the $p$-like Ge L$_{6c}$-derived triplet states, which otherwise remain closely spaced in energy (by $\lesssim 0.1$ eV) with increasing Pb composition $x$. Additionally, we observe that the alloy CB states originating from the $\Gamma_{7c}$ states of Ge decrease in energy more rapidly with increasing $x$ than those originating from the Ge L$_{6c}$ states. As $x$ increases the calculated CB states originating from the $\Gamma_{7c}$ states of Ge firstly pass through the $p$-like Ge L$_{6c}$-derived CB states, for $x \approx 2.5$\%. As this occurs we note weak hybridisation between these two sets of states, evidenced by the abrupt rise and subsequent fall in the values of $\frac{ \textrm{d} E_{g} }{ \textrm{d} P }$ associated with the $p$-like triplet CB states. As $x$ increases further, we note the emergence of a direct band gap for $x \approx 7$\%. This is evidenced by a sharp increase in $\frac{ \textrm{d} E_{g} }{ \textrm{d} P }$ associated with the fundamental band gap as the energy of the Ge $\Gamma_{7c}$-derived CB states approaches that of the $s$-like Ge L$_{6c}$-derived singlet states (cf.~Fig.~\ref{fig:sqs_summary}(d)). This sharp increase in $\frac{ \textrm{d} E_{g} }{ \textrm{d} P }$ describes the transition to a fundamental band gap having primarily, but not purely, direct (Ge $\Gamma_{7c}$) character. We emphasise that this transfer of Ge $\Gamma_{7c}$ character to the Ge L$_{6c}$-derived states again occurs only when the two sets of states become close in energy, suggesting weak Pb-induced hybridisation of Ge L$_{6c}$ and $\Gamma_{7c}$ host matrix states. We note that this is in contrast to the results of equivalent analysis carried out for Ge$_{1-x}$Sn$_{x}$ SQSs, \cite{Broderick_NUSOD_2019} in which we find that the pressure coefficient associated with the fundamental band gap increases monotonically with increasing Sn composition $3 \lesssim x \lesssim 12$\%, indicating an indirect- to direct-gap transition characterised by strong band hybridisation in the supercells considered. \cite{Halloran_OQE_2019,Eales_SR_2019}

The alloy band gap becomes primarily direct in character by $x = 7.81$\% in the 128-atom SQS calculations -- i.e.~for the Ge$_{118}$Pb$_{10}$ SQS -- at which composition the band gap simultaneously closes, suggesting a semiconducting to semimetallic transition that coincides with the indirect- to direct-gap transition. The character of the indirect- to direct-gap transition in Ge$_{1-x}$Pb$_{x}$ is emphasised in Fig.~\ref{fig:sqs_summary}(d), which shows the variation with $x$ of $\frac{ \textrm{d} E_{g} }{ \textrm{d} P }$ associated with the fundamental band gap of Fig.~\ref{fig:sqs_summary}(c). Here, as $x$ increases $\frac{ \textrm{d} E_{g} }{ \textrm{d} P }$ initially remains close to that associated with the fundamental (indirect) L$_{6c}$-$\Gamma_{8v}$ band gap of Ge (cf.~Table~\ref{tab:ordered_supercell_results}). However, as $x$ increases above 6\% we calculate an abrupt increase in $\frac{ \textrm{d} E_{g} }{ \textrm{d} P }$, towards that of the direct $\Gamma_{7c}$-$\Gamma_{8v}$ band gap of Ge (cf.~Table~\ref{tab:ordered_supercell_results}), reflecting the onset of an alloy band gap having primarily direct character over a narrow range of Pb compositions centred about $x \approx 7$\%.


While the calculated values of $\frac{ \textrm{d} E_{g} }{ \textrm{d} P }$ in Fig.~\ref{fig:sqs_summary}(d) illustrate the presence of an indirect- to direct-gap transition in Ge$_{1-x}$Pb$_{x}$, we note that these values are not suitable for direct comparison to experimental measurements, and likewise that Fig.~\ref{fig:sqs_summary}(c) may overestimate the Pb composition at which the indirect- to direct-gap transition may occur. In a real alloy, all states that can (by symmetry) hybridise will hybridise: this behaviour is not captured quantitatively in DFT supercell calculations due to their limitation to structures containing $\lesssim 10^{2}$ atoms. In such small supercells only a limited number of Ge host matrix states fold to $\textbf{K} = 0$ close in energy to the CB edge. Since only states which fold to the same supercell wave vector \textbf{K} can hybridise in supercell calculations, the size of a given alloy supercell then limits the number of states that can hybridise with the $\Gamma$-point eigenstates of the Ge host matrix semiconductor. Quantitative description of the emergence of a direct band gap -- characterised in Ge$_{1-x}$Pb$_{x}$ and the related group-IV alloy Ge$_{1-x}$Sn$_{x}$ by the emergence of Ge $\Gamma_{7c}$ character at the CB edge -- is then limited in small alloy supercell calculations. In practice, this limits the growth of $\frac{ \textrm{d} E_{g} }{ \textrm{d} P }$ with increasing $x$, with calculations for small disordered alloy supercells producing values of $\frac{ \textrm{d} E_{g} }{ \textrm{d} P }$ which can be expected to underestimate experimental measurements.

Additionally, calculations for SQSs containing $\lesssim 10^{2}$ atoms suffer from finite-size effects, insofar as the SQSs possess a higher degree of ordering than in a real disordered alloy. This stems from the use of Born-von Karman (supercell) boundary conditions, which introduce artificial long-range ordering on a length scale defined by the supercell lattice constant $A$ (with $A \sim 1$ nm in the 128-atom SQSs considered here). One impact of this supercell-imposed ordering is, for instance, that the calculated energy gaps in the 64-atom supercells of Fig.~\ref{fig:ordered_supercell_band_structures}(d) and in Figs.~\ref{fig:disordered_supercell_band_structures}(a) --~\ref{fig:disordered_supercell_band_structures}(d) are in general larger than those computed at the same Pb composition in the 128-atom SQSs. In addition, this ordering allows -- despite the presence of nominally random short-range alloy disorder in a SQS -- for the formation of Ge L$_{6c}$-derived alloy CB states having predominantly $s$-like orbital character on each of the Pb lattice sites. This is observed via the large splitting of $\approx 150$ meV observed between the $s$-like singlet and $p$-like triplet Ge L$_{6c}$-derived states in Fig.~\ref{fig:sqs_summary}(c). Such a large splitting, which closely reflects that observed in ordered alloy supercell calculations (cf.~Sec.~\ref{sec:results_ordered}), would not be expected to occur in a real disordered Ge$_{1-x}$Pb$_{x}$ alloy, due to the lower overall symmetry associated with a lack of long-range ordering. Rather, the primarily Ge L$_{6c}$-related CB states will be inhomogeneously broadened in energy due to the loss of both short- and long-range order. If we assume that the Ge L$_{6c}$-derived alloy CB states are inhomogeneously broadened about their average energy obtained from our SQS calculations (cf.~Fig.~\ref{fig:sqs_summary}(c)), then the Ge $\Gamma_{7c}$-derived CB states would pass below this average energy for $x \approx 3$ -- 4\%. This then places an estimated lower limit from our calculations on the Pb composition $x$ at which Ge$_{1-x}$Pb$_{x}$ transitions from indirect- to direct-gap. This uncertainty in transition composition could be reduced by using large-scale alloy supercells containing $\gtrsim 10^{3}$ -- $10^{4}$ atoms, as we have demonstrated for Ge$_{1-x}$Sn$_{x}$ alloys via the use of a semi-empirical tight-binding model. \cite{Broderick_GeSn_preparation_2019} However, while the computed values of $\frac{ \textrm{d} E_{g} }{ \textrm{d} P }$ shown in Figs.~\ref{fig:sqs_summary}(c) and~\ref{fig:sqs_summary}(d) are expected to underestimate those in real Ge$_{1-x}$Pb$_{x}$ alloys at higher Pb compositions, we emphasise that the qualitative trend predicted in Fig.~\ref{fig:sqs_summary}(d) -- i.e.~that the composition range in which the indirect- to direct-gap transition occurs in Ge$_{1-x}$Pb$_{x}$ can be identified via an accompanying sharp increase in $\frac{ \textrm{d} E_{g} }{ \textrm{d} P }$ -- can be expected to emerge in real alloys, and hence in experimental measurements.


Finally, Fig.~\ref{fig:sqs_summary}(e) shows the calculated variation of $\Delta_{\scalebox{0.6}{\textrm{SO}}}$ with $x$. As for the calculation of $E_{g}$ above, $\Delta_{\scalebox{0.6}{\textrm{SO}}}$ is calculated with respect to the average VB edge energy in each SQS. Still, we note deviations from a smooth monotonic increase of $\Delta_{\scalebox{0.6}{\textrm{SO}}}$ with $x$, reflecting (in part) the non-monotonic variation with $x$ of the VB edge splitting associated with short-range alloy disorder. Generally, we find that $\Delta_{\scalebox{0.6}{\textrm{SO}}}$ increases strongly with increasing $x$, although slightly less strongly than the corresponding increase calculated in Sec.~\ref{sec:results_ordered} for ordered alloy supercells. For example, the calculated value $\Delta_{\scalebox{0.6}{\textrm{SO}}} = 0.571$ eV for an ordered Ge$_{15}$Pb$_{1}$ ($x = 6.25$\%) supercell is larger by 45 meV than the value of 0.526 eV calculated for the Ge$_{119}$Pb$_{9}$ ($x = 7.03$\%) SQS, the latter being the largest value calculated for the series of SQSs considered. Nonetheless, the overall increase of $\Delta_{\scalebox{0.6}{\textrm{SO}}}$ with $x$ is sufficiently strong that we find $\Delta_{\scalebox{0.6}{\textrm{SO}}} > E_{g}$ for Pb compositions as low as $x \approx 2.3$\%. We note that this behaviour is qualitatively similar to that in the III-V dilute bismide alloy (In)GaAs$_{1-x}$Bi$_{x}$, where Bi incorporation results in a simultaneous strong decrease of $E_{g}$ and increase of $\Delta_{\scalebox{0.6}{\textrm{SO}}}$ with increasing $x$, leading to a band structure in which $\Delta_{\scalebox{0.6}{\textrm{SO}}} > E_{g}$. This unusual band structure condition is appealing from a practical perspective since it offers the potential to suppress (i) hot-hole producing non-radiative Auger recombination processes, and (ii) inter-valence band absorption involving the spin-split-off band, both of which play a strong role in limiting the efficiency of long-wavelength semiconductor lasers and light-emitting diodes. \cite{Broderick_SST_2012} Our calculations here suggest that direct-gap Ge$_{1-x}$Pb$_{x}$ alloys should also have $\Delta_{\scalebox{0.6}{\textrm{SO}}} > E_{g}$, presenting additional scope for band structure engineering.


\section{Conclusions}
\label{sec:conclusions}


In summary, we have presented a theoretical analysis of electronic structure evolution and the indirect- to direct-gap transition in the group-IV alloy Ge$_{1-x}$Pb$_{x}$. We established DFT calculations -- using both the HSEsol and mBJ XC functionals -- of the structural and electronic properties of the constituent materials (i) diamond-structured semiconducting Ge, (ii) diamond-structured semimetallic Pb, and (iii) the fictitious semimetallic IV-IV zinc blende-structured compound GePb. Using applied hydrostatic pressure as a probe of the electronic structure of Ge$_{1-x}$Pb$_{x}$ alloy supercells, we (i) demonstrated that Pb incorporation primarily impacts the CB structure, and (ii) elucidated the mechanism driving the indirect- to direct-gap transition. Comparison of the HSEsol- and mBJ-calculated electronic structure of ordered alloy supercells established the suitability of the mBJ XC functional to analyse Ge$_{1-x}$Pb$_{x}$ alloys. The mBJ XC functional was therefore used to analyse the electronic structure evolution in disordered and SQS alloy supercells.


Beginning with ordered alloy supercells, we found that Pb incorporation in a Ge$_{63}$Pb$_{1}$ ($x = 1.56$\%) or Ge$_{15}$Pb$_{1}$ ($x = 6.25$\%) supercell partially lifts the degeneracy of the Ge L$_{6c}$ CB edge states, giving rise to (i) a singlet possessing $s$-like symmetry at the Pb lattice site, and (ii) a triplet possessing $p$-like symmetry at the Pb lattice site. The emergence of an $s$-like singlet state at the CB edge could be interpreted as evidence of a direct band gap for Pb compositions as low as $x \approx 1$\%. However, the calculated pressure coefficient associated with the Ge$_{63}$Pb$_{1}$ supercell band gap demonstrated that the CB edge singlet state possesses only a small admixture of direct (Ge $\Gamma_{7c}$) character and retains primarily indirect (Ge L$_{6c}$) character at $x = 1.56$\%. For a Ge$_{15}$Pb$_{1}$ ($x = 6.25$\%) supercell, the band gap pressure coefficient was calculated to be close to that of the Ge direct $\Gamma_{7c}$-$\Gamma_{8v}$ band gap, indicating the emergence of a direct band gap in the alloy with increasing Pb composition $x$, characterised by the transfer of Ge $\Gamma_{7c}$ character to the alloy CB edge and driven by Pb composition-dependent band hybridisation. This supported the requirement for further detailed analysis of the electronic structure evolution in disordered alloy supercells, to quantify the nature of the indirect- to direct-gap transition and therefore identify the Pb composition $x$ at which Ge$_{1-x}$Pb$_{x}$ becomes a direct-gap semiconductor.

To quantify the impact of Pb-related alloy disorder we tracked the evolution of the alloy CB edge in a Ge$_{62}$Pb$_{2}$ supercell as the distance between the Pb atoms was reduced from fourth- to first-nearest neighbour. Substituting two Pb atoms at successively closer lattice sites, we found strong dependence of the alloy band gap on the separation between the two Pb atoms. From a theoretical perspective our calculations therefore indicate that short-range alloy disorder must be explicitly considered to quantitatively analyse the Ge$_{1-x}$Pb$_{x}$ electronic structure, suggesting that VCA-based approaches are ill-suited to analyse Ge$_{1-x}$Pb$_{x}$ alloys. Having established the importance of the alloy microstructure in determining the details of the electronic properties, we then analysed the emergence of a direct band gap with increasing $x$ for the realistic case of a randomly disordered alloy using a SQS approach.

The calculated electronic structure evolution for disordered 128-atom SQSs again showed a singlet state at the CB edge. However, the calculated pressure coefficients associated with the energy gaps between the VB edge and the five lowest energy CB states showed that the CB edge retained primarily Ge L$_{6c}$ character until $x \approx 6$ -- 7\%. In this composition range the lowest energy CB state rapidly acquired predominantly direct (Ge $\Gamma_{7c}$) character, evidenced by an abrupt increase in the fundamental band gap pressure coefficient. The SQS calculations therefore indicate an indirect- to direct-gap transition in Ge$_{1-x}$Pb$_{x}$ alloys for $x \approx 6$ -- 7\%, near which composition the CB edge also passes through the VB edge, to give a semimetallic alloy.

However, finite-size effects in the SQSs considered are expected to lead to an overestimate of the composition at which Ge$_{1-x}$Pb$_{x}$ becomes direct-gap. Our SQS calculations show large energy splitting of the four Ge L$_{6c}$-derived alloy CB states, with the CB edge singlet state (having $s$-like orbital character at Pb lattice sites) lying $\approx 150$ meV below the Ge L$_{6c}$-derived triplet states (having $p$-like orbital character at Pb lattice sites). We expect that such a large splitting would not be observed in larger supercells, or in real alloys, and arises in the SQSs considered due to the absence of long-range alloy disorder in the relatively small supercells to which DFT calculations are limited. Rather, we anticipate that the Ge L$_{6c}$ character associated with alloy CB states will, in real alloys, experience inhomogeneous energetic broadening about the mean energy of alloy CB states possessing Ge L$_{6c}$ character, due to further reduction in symmetry associated with long-range alloy disorder. The CB state possessing greatest Ge $\Gamma_{7c}$ character passes through the average energy of the four Ge$_{6c}$-derived alloy CB states in our SQS calculations for $x \approx 3$ -- 4\%. This average may then provide a more realistic estimate of the composition range in which Ge$_{1-x}$Pb$_{x}$ becomes a direct-gap semiconductor, at which composition the calculated alloy band gap is in the range 0.3 -- 0.4 eV.


Overall, we predict the emergence of a direct band gap in response to substitutional Pb incorporation in Ge, suggesting that group-IV Ge$_{1-x}$Pb$_{x}$ alloys are potentially of interest for applications in CMOS-compatible active photonic devices operating at mid-infrared wavelengths. However, the potential for applications of Ge$_{1-x}$Pb$_{x}$ alloys in such devices may in practice be limited to wavelengths deep in the infrared by the presence of a low fundamental band gap. Further investigations are now required to confirm the emergence of a direct band gap in response to Pb incorporation, as well as to quantify the implications of the alloy band mixing and disorder effects identified by our analysis for technologically relevant material properties (e.g.~optical generation and recombination rates, carrier mobilities, and band-to-band tunneling rates).


\section*{Acknowledgements}

This work was supported by Science Foundation Ireland (SFI; project nos.~15/IA/3082 and 14/IA/2513), and by the National University of Ireland (NUI; via the Post-Doctoral Fellowship in the Sciences, held by C.A.B.). The authors thank Prof.~Justin D.~Holmes and Dr.~Subhajit Biswas (University College Cork, Ireland) for bringing their attention to this topic, and for useful discussions.






\end{document}